\documentclass[trackchanges]{aastex701}
\usepackage{color}
\usepackage{bm}
\usepackage{cases}
\usepackage{multirow}
\usepackage{lineno}
\usepackage{textcomp}

\newcommand{\pccm}{\mathrm{pc~cm^{-3}}}
\newcommand{\kmsMpc}{\mathrm{km~s^{-1}~Mpc^{-1}}}

\begin{document}

\title{Investigating the Anisotropy of Dispersion Measure Contribution from the Galactic Halo by Using Fast Radio Bursts}

\author[orcid=0000-0003-2721-2559]{Yang Liu}
\affiliation{Purple Mountain Observatory, Chinese Academy of Sciences, No. 10 Yuanhua Road, Nanjing 210023, China}
\email[show]{yangliu@pmo.ac.cn}

\author{Bao Wang}
\affiliation{Purple Mountain Observatory, Chinese Academy of Sciences, No. 10 Yuanhua Road, Nanjing 210023, China}
\affiliation{School of Astronomy and Space Sciences, University of Science and Technology of China, No. 96 JinZhai Road, Hefei 230026, China}
\email[show]{baowang@pmo.ac.cn}

\author{Puxun Wu}
\affiliation{Department of Physics and Key Laboratory of Low-Dimensional Quantum Structures and Quantum Control of Ministry of Education, Hunan Normal University, Changsha, Hunan 410081, China}
\affiliation{Hunan Research Center of the Basic Discipline for Quantum Effects and Quantum Technologies, Hunan Normal University, Changsha, Hunan 410081, China}
\email[show]{pxwu@hunnu.edu.cn}

\author{Jun-Jie Wei}
\affiliation{Purple Mountain Observatory, Chinese Academy of Sciences, No. 10 Yuanhua Road, Nanjing 210023, China}
\affiliation{School of Astronomy and Space Sciences, University of Science and Technology of China, No. 96 JinZhai Road, Hefei 230026, China}
\email[show]{jjwei@pmo.ac.cn}

\author{Xue-Feng Wu}
\affiliation{Purple Mountain Observatory, Chinese Academy of Sciences, No. 10 Yuanhua Road, Nanjing 210023, China}
\affiliation{School of Astronomy and Space Sciences, University of Science and Technology of China, No. 96 JinZhai Road, Hefei 230026, China}
\email[show]{xfwu@pmo.ac.cn}

\correspondingauthor{Jun-Jie Wei}

\begin{abstract}

We propose a data-driven approach to reconstruct the all-sky distribution of the dispersion measure contribution from the Galactic halo ($\mathrm{DM_{halo}}$) through a spherical harmonic expansion, enabling an investigation of its possible anisotropies. Based on the NE2001 model and using 92 localized and 574 unlocalized non-repeating fast radio bursts (FRBs) at Galactic latitudes $|b|>15^\circ$, we find a significant dipole anisotropy in $\mathrm{DM_{halo}}$, pointing toward $(l=130^\circ,\, b=+5^\circ)$ with a $1\sigma$ uncertainty of approximately $28^\circ$. 
The $\mathrm{DM_{halo}}$ value in this direction is $63\pm9~\pccm$, exceeding the all-sky mean by about $2.6\sigma$.
This result is not significantly affected by the choice of Galactic ISM models.
Furthermore, even when using a refined sample of 62 localized FRBs (excluding CHIME detections, repeaters, and unlocalized events), the dipole anisotropic structure persists, with a direction of $(l=141^\circ,\, b=+51^\circ)$ and a larger 1$\sigma$ uncertainty of $\sim 44^\circ$.
Model comparisons using the Akaike Information Criterion and Bayesian evidence yield consistent preferences, and together they suggest that current FRB data slightly favor the existence of a dipole structure in $\mathrm{DM_{halo}}$. 
If this feature is not a statistical fluctuation or systematic error, its physical origin requires further investigation. Future FRB samples with larger sizes and more complete sky coverage will be essential to confirm or refute this possible anisotropic structure.

\end{abstract}
			

\section{Introduction} 
Galaxies like the Milky Way (MW) are enveloped by an extensive halo that reaches approximately to the virial radius. The gas in the halo (also referred to as the circumgalactic medium, CGM) is complex and multiphase, with temperatures ranging from $\sim 10^4$--$10^7~\mathrm{K}$. It represents a significant reservoir of the Galaxy's baryons and serves as a key interface between the Galactic disk and the intergalactic medium (IGM) \citep{2012ARA&A..50..491P,2017ARA&A..55..389T}.
Detecting the physical properties of the halo gas, such as its distribution, spatial extent, and mass, is essential for understanding galaxy formation, accretion, and feedback mechanisms.
Because most halo gas is highly diffuse and hot, however, its direct detection is difficult.
Currently, the primary method for probing the MW halo relies on X-ray observations, including diffuse emission lines of O {\footnotesize VII} and O {\footnotesize VIII} and absorption lines in the spectra of active galactic nuclei (AGN).
Such observations have been widely employed to estimate the mass, structure, and other properties of the halo gas~\citep{2007PASJ...59S...9S,2013ApJ...773...92H,2005ApJ...635..386W,2007ApJ...669..990B,2013ApJ...762...20F,2015ApJS..217...21F, 2010ApJ...714..320A,2012ApJ...756L...8G,2018ApJ...862....3B,2013ApJ...770..118M,2015ApJ...800...14M,2017ApJ...849..105L,2018ApJ...862...34N,2020NatAs...4.1072K,2020ApJ...888..105Y,2020MNRAS.496L.106K}.
An alternative and complementary observable is the dispersion measure (DM) of radio signals, which encodes the integrated column density of free electrons along the line of sight and thus provides information on the ionized baryon content. Observations of pulsars in the Large and Small Magellanic Clouds (LMC and SMC) can be used to estimate the DM contribution from the Galactic halo ($\mathrm{DM_{halo}}$) in those specific directions~\citep{2013MNRAS.433..138R}. However, each of these methods has limitations. 
X-ray emission and absorption observations often depend on assumptions about the gas metallicity and ionization state, which can lead to substantially different results.
Furthermore, X-ray absorption line data are limited to a small number of sightlines, preventing the reconstruction of the all-sky halo gas distribution.
Measurements using LMC and SMC pulsars are also limited, as these galaxies are at a distance of only $\sim 50~\mathrm{kpc}$--$60~\mathrm{kpc}$; thus, they probe only the inner halo and not its full extent to the virial radius.
Therefore, a probe that is both abundant and provides direct column density measurements would be highly valuable for investigating the distribution and properties of the Galactic halo gas.

Fast radio bursts (FRBs) are bright, millisecond-duration transients that offer a powerful alternative probe of baryonic matter through the measurement of their DMs. 
The first FRB was discovered in archival data from the Parkes Telescope~\citep{2007Sci...318..777L}.
Unlike pulsars in the Magellanic Clouds, which probe only nearby regions, FRBs are detected at cosmological distances and thus have become a widely used tool in cosmological studies~\citep{2019A&ARv..27....4P, 2022A&ARv..30....2P, 2026enap....5..448G, 2024ChPhL..41k9801W, 2025PASA...42...17H}.
Owing to their high all-sky event rate ($\sim10^3~\mathrm{day}^{-1}$; \citealt{2018MNRAS.475.1427B}) and the fact that their observed $\mathrm{DM_{obs}}$ includes contributions from intervening galactic halos along the sightlines as well as from the Galactic halo out to its virial radius, FRBs provide a powerful tool for investigating the halo gas in intervening galaxies or in the MW.
A recent study by \cite{2019MNRAS.485..648P} indicated that M31's halo can be easily detected by large FRB surveys.
\cite{2019ApJ...872...88R} also showed that the baryon fractions in the CGM of intervening galaxies along the FRBs sightlines and in the IGM can be accurately estimated with only a few tens of FRBs at $z<1$, using the simulated FRBs.
Later, \cite{2022NatAs...6.1035C} analyzed a sample of 474 distant FRBs and found that the mean DM of galaxy-intersecting FRBs was larger than that of non-intersecting FRBs with $>99\%$ confidence. 
Similar results were reported by \cite{2023ApJ...945...87W}, who detected a DM excess from the CGM of $10^{11}$--$10^{13}~M_\odot$ halos at $1$--$2\sigma$ significance.
Most recently, \cite{2025NatAs...9.1226C} reported evidence for efficient feedback processes that can expel gas from galaxy halos into the IGM.
These studies indicate that FRBs provide a powerful probe of the gaseous halos of external galaxies and their baryonic content.

For the halo gas in the MW, on the other hand,  \cite{2020ApJ...895L..49P} combined the pulsar and FRB data and estimated that $\mathrm{DM_{halo}}$ lies between $-2$ and $123~\pccm$ at the 95\% confidence level (CL). \cite{2023ApJ...946...58C} analyzed FRB data at Galactic latitudes $|b|>30^\circ$ and obtained upper limits on $\mathrm{DM_{halo}}$, ranging from $52$ to $111~\pccm$. Recently, \cite{2025AJ....169..330R} analyzed FRB 20220319D together with two nearby pulsars in the sky and derived upper limits on $\mathrm{DM_{halo}}$ of $28.7~\pccm$ and $47.3~\pccm$, respectively.

It should be noted that previous studies either provided only upper limits on $\mathrm{DM_{halo}}$ across the entire sky or along specific sightlines, or reported broad ranges. The directional dependence of $\mathrm{DM_{halo}}$ was not taken into account.
However, the MW's CGM may be anisotropic. It contains known structures that extend across several to tens of kpc, such as the Fermi bubbles~\citep{2010ApJ...724.1044S}, polarized radio lobes~\citep{2013Natur.493...66C}, and the eROSITA bubbles~\citep{2020Natur.588..227P}.
Recently, using X-ray emission measure (EM) along 107 sightlines of hot halo gas, \cite{2018ApJ...862...34N} demonstrated that the observed all-sky EM distribution cannot be explained by a spherically symmetric electron density model.
\cite{2020ApJ...888..105Y} later proposed a combined spherical and disk-like gas component model (the YT2020 model), which reproduces the directional dependence of the X-ray EM data and predicts direction-dependent $\mathrm{DM_{halo}}$.
Simulations from the High-resolution Environmental Simulations of the Immediate Area (H{\footnotesize ESTIA}) also showed that most $\mathrm{DM_{halo}}$ fluctuations occur on angular scales of $\gtrsim 10^\circ$, with values ranging from $13~\pccm$ to $166~\pccm$ and a mean of $45~\pccm$~\citep{2025MNRAS.538.2785H}.
These studies imply that $\mathrm{DM_{halo}}$ may vary significantly across different sightlines.

Therefore, to test whether the DM contribution from the Galactic halo exhibits anisotropy, we propose a data-driven approach to reconstruct the all-sky distribution of $\mathrm{DM_{halo}}$. Specifically, we expand $\mathrm{DM_{halo}}$ in spherical harmonics and construct models of varying expansion degrees, enabling us to capture possible anisotropic structures. 
We constrain these models using a set of 92 localized FRBs and 574 unlocalized non-repeating FRBs. By comparing isotropic and anisotropic $\mathrm{DM_{halo}}$ models, we determine which is statistically preferred, thereby assessing the presence of anisotropy in $\mathrm{DM_{halo}}$.

This paper is organized as follows: In Section \ref{sec:method}, we introduce the method used in our analysis, including the spherical harmonic expansion and the construction of the likelihood function for FRBs. The sample selection and results are shown in Section~\ref{sec:samples_results}, and the conclusions are summarized in Section~\ref{sec:conclusion}.

\section{Methodology}\label{sec:method}
\subsection{$\mathrm{DM_{halo}}$ Model Expanded by Spherical Harmonics}
For an extragalactic FRB, its DM can be divided into several components:
\begin{eqnarray}\label{eq:DMobs}
	\mathrm{DM_{obs}}=\mathrm{DM_{MW}}+\mathrm{DM_{ext}},
\end{eqnarray}
where $ \mathrm{DM_{MW}} = \mathrm{DM_{ISM}}+\mathrm{DM_{halo}} $ represents the contribution from the medium within the MW, and $ \mathrm{DM_{ext}}=\mathrm{DM_{cos}}(z)+\mathrm{DM_{host}}(z) $ represents the contribution from the extragalactic medium. The subscripts ``ISM'', ``halo'', ``host'', and ``cos'' denote the contributions from the ionized medium in the MW interstellar medium (ISM), the Galactic halo gas, the FRB host galaxy, and the IGM together with intervening halos, respectively.
For the sake of using FRBs to investigate the all-sky distribution of $\mathrm{DM_{halo}}$, we treat $\mathrm{DM_{halo}}$ as a function varying with sightlines and expand it using spherical harmonics. 

Spherical harmonics arise as the angular part of the solutions to Laplace's equation in spherical coordinates. They form a complete set of orthonormal basis functions on the sphere and can be used to approximate functions defined on the spherical surface.
In the usual spherical coordinates $(\theta,\phi)$, with $\theta$ being the polar angle ($0\leq \theta \leq \pi$) and $\phi$ being the azimuthal angle ($0\leq \phi < 2\pi$), the complex spherical harmonics are defined as
\begin{equation}
	Y_{\ell, m}(\theta,\phi) =
	N_{\ell, m}\,P_{\ell}^{m}(\cos\theta)\,e^{i m \phi},
\end{equation}
where the degree $\ell$ is a non-negative integer, and the order $m$ is an integer satisfying $-\ell \leq m \leq \ell$. $P_{\ell}^m$ are the associated Legendre functions and 
\begin{eqnarray}
	N_{\ell, m} = \sqrt{\frac{2\ell+1}{4\pi}\,\frac{(\ell-m)!}{(\ell+m)!}}
\end{eqnarray}
is the normalization constant.
The real spherical harmonics can be obtained as linear combinations of the complex ones:
\begin{eqnarray}\label{eq:Ylm}
	Y^{\mathrm{R}}_{\ell, m}(\theta,\phi) =
	\left\{\begin{array}{ll}
		\frac{1}{\sqrt{2}}\left[Y_{\ell, m}(\theta,\phi) + (-1)^m Y_{\ell,-m}(\theta,\phi)\right], & m>0, \\
		Y_{\ell, 0}(\theta,\phi), & m=0, \\
		\frac{1}{i\sqrt{2}}\left[Y_{\ell, |m|}(\theta,\phi) - (-1)^{|m|}Y_{\ell,-|m|}(\theta,\phi)\right], & m<0,
	\end{array} \right.
\end{eqnarray}
Then, the function $\mathrm{DM_{halo}}$ can be expanded in terms of these real spherical harmonics $Y^{\mathrm{R}}_{\ell, m}(\theta,\phi)$:
\begin{eqnarray} \label{eq:dm_harmonic_expansion}
	\mathrm{DM_{halo}}(\theta,\phi)
	= \sum_{\ell=0}^{\ell_{\max}}\sum_{m=-\ell}^{\ell} a_{\ell, m}\;Y^{\mathrm{R}}_{\ell, m}\left(\theta,\phi\right),  
\end{eqnarray}
with
\begin{equation}
	\label{eq:spherical_harmonics}
	\begin{array}{r@{\,}c@{\,}l@{\qquad}r@{\,}c@{\,}l@{\qquad}r@{\,}c@{\,}l}
		Y^\mathrm{R}_{0,0} & = & \frac{1}{2\sqrt{\pi}}, &
		Y^\mathrm{R}_{1,0} & = & \frac{1}{2}\sqrt{\frac{3}{\pi}}\cos(\theta), &
		Y^\mathrm{R}_{1,1} & = & -\frac{1}{2}\sqrt{\frac{3}{\pi}} \cos(\phi)\sin(\theta), \\[2ex]
		Y^\mathrm{R}_{1,-1} & = & -\frac{1}{2}\sqrt{\frac{3}{\pi}} \sin(\phi) \sin(\theta), &
		Y^\mathrm{R}_{2,0} & = & \frac{1}{4}\sqrt{\frac{5}{\pi}}(3\cos^2(\theta)-1), &
		Y^\mathrm{R}_{2,1} & = & -\frac{1}{4}\sqrt{\frac{15}{\pi}}\cos(\phi)\sin(2\theta), \\[2ex]
		Y^\mathrm{R}_{2,-1} & = & -\frac{1}{4}\sqrt{\frac{15}{\pi}} \sin(\phi) \sin(2\theta), &
		Y^\mathrm{R}_{2,2} & = & \frac{1}{4}\sqrt{\frac{15}{\pi}} \cos(2\phi)\sin^2(\theta), &
		Y^\mathrm{R}_{2,-2} & = & \frac{1}{4}\sqrt{\frac{15}{\pi}} \sin(2\phi) \sin^2(\theta), \quad \cdots.
	\end{array}
\end{equation}
Here, we can use the relation  
$\theta=\frac{\pi}{2}-b,\ \text{and}\ \phi=l, $
to convert the spherical coordinates $(\theta,\phi)$ into the commonly used Galactic coordinates $(l,\,b)$, where $(0\leq l\leq 2\pi)$ is the Galactic longitude and $\left(-\frac{\pi}{2}\leq b\leq \frac{\pi}{2}\right)$ is the Galactic latitude. 
In Equation~(\ref{eq:dm_harmonic_expansion}), the monopole $(\ell=0)$ gives the all-sky mean $\mathrm{DM_{halo}}$ contribution, the dipole $(\ell=1)$ captures directional excess or deficit of $\mathrm{DM_{halo}}$, and higher $\ell$ describes progressively finer structure.
Therefore, using the FRB data, we can estimate the expansion coefficients $a_{\ell,m}$ and thus quantify the variation of $\mathrm{DM_{halo}}$ across the sky without assuming any density profile models.

\subsection{Likelihood Function of FRBs}

To obtain the all-sky $\mathrm{DM_{halo}}$ distribution from FRB data, we follow the common strategy in FRB cosmology, employing a Bayesian statistical framework to estimate the spherical harmonic coefficients $a_{\ell, m}$ in Equation~(\ref{eq:dm_harmonic_expansion}). 
This process relies on constructing a likelihood function for the FRB data, which is derived from the probability density functions (PDFs) for $\mathrm{DM_{ISM}}$, $\mathrm{DM_{host}}$ and $\mathrm{DM_{cos}}$.

The contribution from the ISM, denoted as $\mathrm{DM_{ISM}}$, can be estimated using either the NE2001~\citep{2002astro.ph..7156C} or YMW16 model~\citep{2017ApJ...835...29Y}. Since their estimates are generally considered accurate to within a factor of two~\citep{2012MNRAS.427..664S}, we model the PDF of the ISM contribution using a truncated Gaussian distribution with a standard deviation of $\sigma_\mathrm{ISM}=\overline{\mathrm{DM}}_\mathrm{ISM}/2$:
\begin{eqnarray}
		P_{\mathrm{ISM}}(\mathrm{DM_{ISM}}) \propto 
	\left\{ 
	\begin{array}{ll} 
		\frac{1}{\sqrt{2\pi}\sigma_\mathrm{ISM}}\exp\left[ -\frac{\left( \mathrm{DM_{ISM}} - \overline{\mathrm{DM}}_\mathrm{ISM}  \right)^2}{2\sigma_\mathrm{ISM}^2} \right] ,&\  \text{if}\ 0\leq \mathrm{DM_{ISM}}\leq \mathrm{DM_{obs}}, \\
		0,&\ \text{else},
	\end{array}
	\right.
\end{eqnarray}
where $\overline{\mathrm{DM}}_\mathrm{ISM}$ is the mean value derived from the ISM model.

The contribution from an FRB's host galaxy, $\mathrm{DM_{host}}$, is commonly modeled using a log-normal distribution~\citep{2020Natur.581..391M,2020ApJ...900..170Z}, which is characterized by the parameters $\mu_\mathrm{host}$ and $\sigma_{\mathrm{host}}$:
\begin{eqnarray}\label{eq:P_host}
	P_{\mathrm{host}}(\mathrm{DM}_{\mathrm{host}})=\frac{1}{\sqrt{2\pi}\mathrm{DM}_{\mathrm{host}}\sigma_{\mathrm{host}}}\exp\left[-\frac{(\ln\mathrm{DM}_{\mathrm{host}}-\mu_{\mathrm{host}})^{2}}{2\sigma_{\mathrm{host}}^{2}}\right].
\end{eqnarray}
These parameters vary with both redshift and host galaxy type. 
For our analysis, we adopt the values obtained by \cite{2020ApJ...900..170Z} from the IllustrisTNG cosmological simulation. 
Their work provides the best-fit parameters for $\mu_\mathrm{host}$ and $\sigma_{\mathrm{host}}$ for three distinct types of FRB host galaxies at eight redshift points between $z=0.1$ and $z=1.5$. 
Specifically, repeaters like FRB 121102 localized to dwarf galaxies are often classified as Type~I; repeaters like FRB 20180916 localized to spiral galaxies are classified as Type~II; and non-repeating FRBs are classified as Type~III. The recent data catalogs containing these FRB host galaxy types are presented in \citep{2025A&A...698A.215G, 2025arXiv250706841Z, 2025ApJ...988..177X, 2025ApJ...981....9W}.
To determine the PDF of $\mathrm{DM_{host}}$ for each FRB in our sample, we assign each FRB to one of these three host galaxy types and obtain the corresponding $\mu_\mathrm{host}$ and $\sigma_{\mathrm{host}}$ at its redshift by applying cubic spline interpolation.
It is important to note that the results from \cite{2020ApJ...900..170Z} do not account for the DM contributed by the circumprogenitor medium due to its high uncertainty. Consequently, in the subsequent analysis, we exclude sources expected to have significant DM contributions from this component.

The $\mathrm{DM_{cos}}$ is contributed by the IGM together with intervening halos along the FRB propagation path and is generally the dominant component in $\mathrm{DM_{obs}}$. Since the distribution of free electrons along the line of sight is highly inhomogeneous, the measured $\mathrm{DM_{cos}}$ values fluctuate around their average value, $\langle\mathrm{DM_{cos}}\rangle$.
The PDF of $\mathrm{DM_{cos}}$ can be assumed as~\citep{2020Natur.581..391M,2014ApJ...780L..33M,2025arXiv250406845Z,2026ApJ...996...66Z}:
\begin{eqnarray} \label{eq:P_cos}
	P_{\mathrm{cos}}(\mathrm{DM_{cos}}) = \frac{1}{\langle\mathrm{DM_{cos}}\rangle} P_{\mathrm{\Delta}}\left(\frac{\mathrm{DM_{cos}}}{\langle\mathrm{DM_{cos}}\rangle}\right),
\end{eqnarray}
where the distribution of the ratio $\Delta \equiv \mathrm{DM_{cos}}/\langle\mathrm{DM_{cos}}\rangle$ is given by
\begin{eqnarray}\label{eq:P_Delta}
	P_{\mathrm{\Delta}}(\Delta)=A\Delta^{-\beta}\exp\left[-\frac{(\Delta^{-\alpha}-C_0)^2}{2\alpha^2\sigma_{d}^2}\right],\quad\Delta>0.
\end{eqnarray}
Here we set $\alpha=\beta=3$, $A$ is a normalization factor, and $C_0$ is chosen such that the mean of $\Delta$ satisfies $\langle \Delta \rangle = 1$. The parameter $\sigma_d$ represents the effective standard deviation of $\mathrm{DM_{cos}}$, commonly parameterized as $\sigma_d = F z^{-0.5}$, with $F$ characterizing the strength of baryon feedback. Following the recent estimate in \citep{2024ApJ...965...57B}, we adopt $F=10^{-0.75}$ in our analysis.

Since FRBs are detected at cosmological distances, the component $\mathrm{DM_{cos}}$ is sensitive to the underlying cosmological model. Its mean value as a function of redshift can be written, for a spatially flat $\Lambda$CDM universe, as~\citep{2003ApJ...598L..79I,2004MNRAS.348..999I,2014ApJ...783L..35D,2020Natur.581..391M}
\begin{eqnarray}\label{eq:DMcos}
	\langle\mathrm{DM}_{\mathrm{cos}}\rangle(z)=\frac{3c\ \Omega_\mathrm{b0}h^2 (100~\kmsMpc)^2}{8\pi G m_pH_0}\int_{0}^{z}\frac{(1 + z')\ f_{d}\ \chi_{e}(z')}{\sqrt{\Omega_\mathrm{m0}(1 + z)^{3}+ \left(1 - \Omega_\mathrm{m0}\right) }}\mathrm{d}z',
\end{eqnarray}
where $c$, $G$, $m_p$, $\Omega_\mathrm{m0}$, and $H_0$ are the speed of light, the gravitational constant, the proton mass, the present matter density parameter, and the Hubble constant, respectively. The parameter $\Omega_\mathrm{b0}h^2$ with $h\equiv H_0/(100\ \kmsMpc)$ gives the physical baryon density today.
The $f_{d}$ denotes the fraction of baryons in the diffuse ionized gas, while $\chi_e(z)$ is the number of free electrons per baryon, $\chi_e(z)=Y_\mathrm{H} \chi_{e,\mathrm{H}}(z)+\frac{1}{2}Y_\mathrm{He}\chi_{e,\mathrm{He}}(z)$, with $Y_\mathrm{H}\simeq 3/4$ and $Y_\mathrm{He}\simeq 1/4$ being the hydrogen and helium mass fractions. Here $\chi_{e,\mathrm{H}}$ and $\chi_{e,\mathrm{He}}$ are the ionization fractions of hydrogen and helium, respectively. 
Since both hydrogen and helium are fully ionized at $z<3$~\citep{2009RvMP...81.1405M,2011MNRAS.410.1096B}, we take $\chi_{e,\mathrm{H}}=\chi_{e,\mathrm{He}}=1$, yielding $\chi_e=7/8$.

For localized FRBs, which have redshift information enabling the calculation of $P_\mathrm{cos}$ and $P_\mathrm{host}$, the joint likelihood function is:
\begin{eqnarray}\label{eq:local_L}
	\mathcal{L}_\mathrm{FRB}^\mathrm{loc}(\mathrm{DM_{obs}}|\bm{\Theta})=\prod_{i=1}^{N_\mathrm{FRB}}  P_i(\mathrm{DM_{obs}}_{,i}|\bm{\Theta}),
\end{eqnarray}
where $N_\mathrm{FRB}$ is the total number of localized FRBs, $\bm{\Theta}$ represents a set of parameters to be fitted, $\mathrm{DM}_{\mathrm{obs},i}$ denotes the observed DM of the $i$-th localized FRB, and $P_i$ is expressed as:
\begin{eqnarray}\label{eq:Pi}
	P_i(\mathrm{DM_{obs}}_{,i}|\bm{\Theta})&=& \int_{0}^{\mathrm{DM}_{\mathrm{exhalo},i} } \int_{0}^{ \mathrm{DM}_{\mathrm{exhalo},i}-\mathrm{DM}_{\mathrm{ISM}} } 
	P_{\mathrm{host}}(\mathrm{DM}_{\mathrm{host}}) \ P_{\mathrm{ISM}}(\mathrm{DM}_{\mathrm{ISM}}) \nonumber \\
	&\times& P_{\mathrm{cos}}( \mathrm{DM}_{\mathrm{exhalo},i}-\mathrm{DM}_{\mathrm{ISM}} -\mathrm{DM}_{\mathrm{host}})
	\ \mathrm{d}\mathrm{DM}_{\mathrm{host}}\ \mathrm{d}\mathrm{DM}_{\mathrm{ISM}},
\end{eqnarray}
where $\mathrm{DM}_{\mathrm{exhalo},i}\equiv\mathrm{DM}_{\mathrm{obs},i}-\mathrm{DM}_{\mathrm{halo}}(l_i,b_i)$.
Because the Galactic halo contribution $\mathrm{DM_{halo}}(l_i,b_i)$ is calculated for each FRB direction through the spherical harmonic expansion (Equation~(\ref{eq:dm_harmonic_expansion})), this likelihood function enables us to estimate the values of spherical harmonic coefficients $a_{\ell, m}$.
It is worth noting that there are currently only slightly over a hundred localized FRBs, while approximately a thousand available FRB sources have been released. Although these unlocalized FRB sources cannot be used to estimate the value of $\mathrm{DM_{halo}}$, they can still provide constraints in the form of upper limits on $\mathrm{DM_{halo}}$.

To incorporate unlocalized FRBs into our analysis, we modify the likelihood function of localized FRBs (Equation~(\ref{eq:local_L})) by applying a weighting term:
\begin{eqnarray}\label{eq:L}
	\mathcal{L}_\mathrm{FRB}(\mathrm{DM_{obs}}|\bm{\Theta}) = \mathcal{L}_\mathrm{FRB}^\mathrm{loc}(\mathrm{DM_{obs}}|\bm{\Theta}) \times \prod_{k=1}^{N_\mathrm{FRB}^\mathrm{unloc}} w(\mathrm{DM}_{\mathrm{ext},k}^\mathrm{unloc}),
\end{eqnarray}
where $N_\mathrm{FRB}^\mathrm{unloc}$ is the number of unlocalized FRBs. The term $\mathrm{DM}_{\mathrm{ext},k}^\mathrm{unloc}$, defined as $\mathrm{DM}_{\mathrm{obs},k}^\mathrm{unloc}-\overline{\mathrm{DM}}_{\mathrm{ISM},k}^\mathrm{unloc}-\mathrm{DM}_\mathrm{halo}(l_k,b_k)$, represents the extragalactic DM component for the $k$-th unlocalized source.
The weighting function is given by:
\begin{eqnarray}\label{eq:ww}
	w(\mathrm{DM_{ext}^{unloc}})= \frac{1}{2} \mathrm{erfc}\left( -\frac{1}{\sqrt{2}} \frac{\mathrm{DM}_{\mathrm{ext}}^\mathrm{unloc}}{\sigma_\mathrm{ISM}^\mathrm{unloc}} \right),
\end{eqnarray}
where `erfc' is the complementary error function.
This weighting function is the form of a Gaussian cumulative distribution function with zero mean and standard deviation $\sigma_\mathrm{ISM}^\mathrm{unloc}\equiv\overline{\mathrm{DM}}_\mathrm{ISM}^\mathrm{unloc}/2$. It serves to penalize unphysical negative $\mathrm{DM_{ext}^{unloc}}$ values arising from model uncertainties. Specifically, the weight $w$ tends to unity when $\mathrm{DM_{ext}^{unloc}}\gg0$ and approaches zero as $\mathrm{DM_{ext}^{unloc}}$ becomes significantly negative\footnote{
We do not use Equation (\ref{eq:ww}) to penalize localized FRBs, since for the localized sample, $\mathrm{DM}_{\mathrm{obs}}$ is consistently greater than $\overline{\mathrm{DM}}_\mathrm{ISM}$ in our analysis, and the integration limits in Equation (\ref{eq:Pi}) do not allow $\mathrm{DM}_{\mathrm{obs}}-\mathrm{DM}_{\mathrm{ISM}}-\mathrm{DM}_{\mathrm{halo}}$ to be less than zero.}.
Then, the posterior PDF of model parameters can be obtained by using Bayes' theorem:
\begin{eqnarray}\label{eq:post_L}
	\mathcal{P}(\bm{\Theta}|\mathrm{DM_{obs}}) \propto \mathcal{L}_\mathrm{FRB}(\mathrm{DM_{obs}}|\bm{\Theta})\times \Pi(\bm{\Theta}),
\end{eqnarray}
where $\Pi(\bm{\Theta})$ denotes the prior PDF of the parameters.

\section{Sample Selection and Results}\label{sec:samples_results}
\subsection{Constraining $\mathrm{DM_{halo}}$ Models using Full Sample}

We use a dataset consisting of 115 localized FRBs and 720 unlocalized non-repeating FRBs. The localized events were compiled by \cite{2025A&A...698A.215G}, while the unlocalized ones are drawn from the database platform \texttt{Blinkverse}~\citep{2023Univ....9..330X}, which collects FRBs detected by various observatories, including FAST, CHIME, GBT, Arecibo, and so on.
To avoid the impact of the complex medium near the Galactic disk, we exclude data with Galactic latitude $|b|<15^\circ$. 
FRB 20190520B, FRB 20210117A, and FRB 20220831A are also removed due to their extreme $\mathrm{DM_{host}}$ values~\citep{2022Natur.606..873N,2023ApJ...948...67B,2025NatAs...9.1226C}, while FRB 20221027A is excluded because it has more than one  candidate host galaxy~\citep{2024Natur.635...61S}.
Then, utilizing the remaining dataset of 92 localized and 574 unlocalized non-repeating FRBs (hereafter referred to as the `full sample'), we calculate the posterior PDF via Equation~(\ref{eq:post_L}), where $\overline{\mathrm{DM}}_\mathrm{ISM}$ values are derived from the NE2001~\citep{2002astro.ph..7156C} and YMW16~\citep{2017ApJ...835...29Y} models.

We consider three $\mathrm{DM_{halo}}$ models, obtained by truncating the spherical harmonic expansion (Equation~(\ref{eq:dm_harmonic_expansion})) at different maximum degrees $\ell_{\max}$. The case $\ell_{\max}=0$ corresponds to the conventional isotropic $\mathrm{DM_{halo}}$ model, $\ell_{\max}=1$ introduces a dipole structure to describe directional excess or deficit in $\mathrm{DM_{halo}}$, and $\ell_{\max}=2$ further includes a quadrupole component. In the analysis, the spherical harmonic coefficients $a_{\ell, m}$ are treated as free parameters constrained by the FRB data. The uniform prior $\mathcal{U}(10, 500)$ is assigned to $a_{0,0}$, and $\mathcal{U}(-150,150)$ to other spherical harmonic coefficients. 
The cosmological parameters are assigned Gaussian priors, obtained from the Planck 2018 results: $\Omega_\mathrm{b0} h^2=0.02237\pm0.00015$, $\Omega_\mathrm{m0}=0.315\pm0.007$, $H_0=67.36\pm0.54~\kmsMpc$~\citep{2020A&A...641A...6P}.
The baryon fraction in the diffuse ionized gas, $f_d$ in Equation~(\ref{eq:DMcos}), is treated as a free parameter with a uniform prior $\mathcal{U}(0.5,0.99)$. It is inferred simultaneously with $a_{\ell, m}$ and will be marginalized in the final results.

{%
	\renewcommand{\arraystretch}{1.5}%
	\begin{deluxetable*}{l|ccccccccc|c|c|c}[tbp]
		\tablecaption{Constraints on $a_{\ell,m}$ of different $\mathrm{DM_{halo}}$ models using the full sample  \label{tab:1}}
		\tablewidth{0pt}
		\tablehead{
			Model & $a_{0,0}$ & $a_{1,0}$ & $a_{1,1}$ & $a_{1,-1}$ & $a_{2,0}$ & $a_{2,1}$ & $a_{2,-1}$ & $a_{2,2}$ & $a_{2,-2}$ & $-2\ln\mathcal{L}_{\max}$ & AIC\,($\ln \mathcal{Z}$) & $\Delta\mathrm{AIC}\,(\ln \mathcal{B})$
		}
		\startdata
		\hline
		 \multicolumn{13}{c}{ 92 localized and 574 unlocalized FRBs (NE2001) } \\
		\hline
		$\ell_{\max}=0$ & $122^{+25}_{-22}$ & - & - & - & - & - & - & - &  - & 1200 & 1210 ($-607.7$) & 0 (0)  \\
		$\ell_{\max}=1$ & $126^{+17}_{-17}$ & $5^{+18}_{-18}$ & $35^{+18}_{-15}$ & $-42^{+14}_{-18}$ & - & - & - & - &  - & 1184 & 1200 ($-605.7$) & $-10$ ($-2$) \\
		$\ell_{\max}=2$ & $142^{+17}_{-17}$ & $-10^{+17}_{-17}$ & $20^{+13}_{-16}$ & $-31^{+18}_{-17}$ & $52^{+20}_{-16}$ & $-3^{+14}_{-14}$ & $-3^{+16}_{-15}$ & $-13^{+14}_{-14}$ & $-2^{+11}_{-13}$ & 1174 & 1200 ($-611.1$) & $-10$ ($3.4$) \\
		\hline
		\hline
		\multicolumn{13}{c}{ 92 localized and 574 unlocalized FRBs (YMW16)} \\
		\hline
		$\ell_{\max}=0$ & $135^{+23}_{-23}$ & - & - & - & - & - & - & - &  - & 1195 & 1205 ($-606.1$) & 0 (0)  \\
		$\ell_{\max}=1$ & $133^{+17}_{-17}$ & $3^{+20}_{-19}$ & $20^{+18}_{-18}$ & $-53^{+12}_{-17}$ & - & - & - & - &  - & 1182 & 1198 ($-604.2$) & $-7$ ($-1.9$) \\
		$\ell_{\max}=2$ & $151^{+16}_{-16}$ & $-14^{+16}_{-16}$ & $9^{+12}_{-14}$ & $-40^{+15}_{-17}$ & $61^{+18}_{-15}$ & $-2^{+12}_{-14}$ & $0^{+16}_{-14}$ & $-18^{+13}_{-13}$ & $-1^{+10}_{-12}$ & 1165 & 1191 ($-607.9$) & $-14$ ($1.8$) \\
		\enddata
		\tablecomments{All results are given as mean values with 1$\sigma$ uncertainties. 
        Here $-2\ln\mathcal{L}_{\max}$ represents $-2$ times the maximum log-likelihood value. 
        $\Delta \mathrm{AIC=AIC-AIC_{ref}}$, and we set the model with $\ell_{\max}=0$ as the reference one. $\ln\mathcal{Z}$ is the logarithmic Bayesian evidence and $\ln \mathcal{B}= \ln\mathcal{Z}_0-\ln\mathcal{Z}_1$, where $\ln\mathcal{Z}_0$ and $\ln\mathcal{Z}_1$ are the logarithmic Bayesian evidences for the reference isotropic $\mathrm{DM_{halo}}$ model ($\ell_{\max}=0$) and a competing alternative model, respectively.
		}
	\end{deluxetable*}
}


\begin{figure}[htbp]
	\centering
	\includegraphics[width=0.5\textwidth]{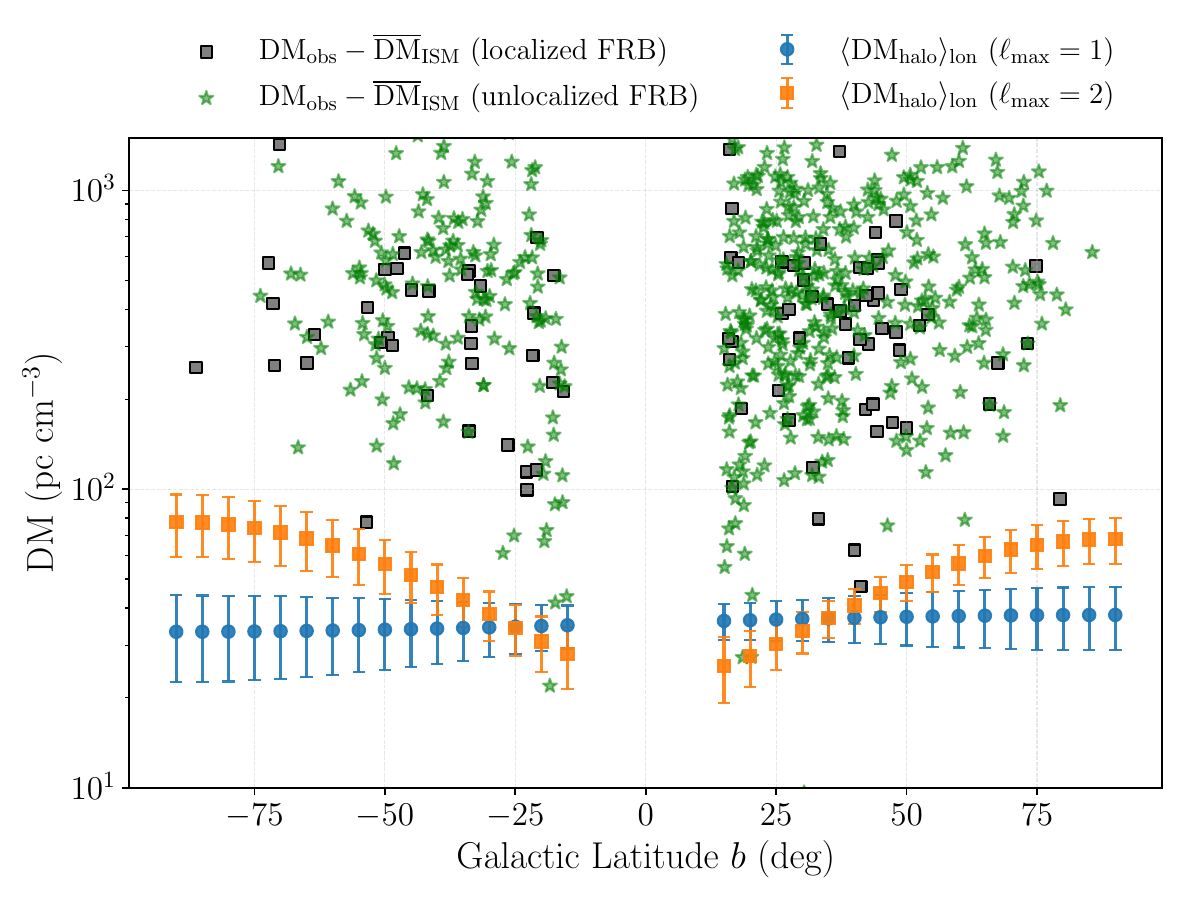}
	\caption{
		Longitude-averaged $\mathrm{DM_{halo}}$ at different Galactic latitudes. Blue and orange points with $1\sigma$ error bars show $\langle\mathrm{DM_{halo}}\rangle_\mathrm{lon}$ for the $\ell_{\max}=1$ and $\ell_{\max}=2$ models, respectively. Gray squares denote localized FRBs, and green stars denote unlocalized FRBs, where $\overline{\mathrm{DM}}_\mathrm{ISM}$ is calculated from the NE2001 model.
		\label{fig:DM_lat}}
\end{figure}

\begin{figure}[htbp]
	\centering
	\includegraphics[width=0.45\textwidth]{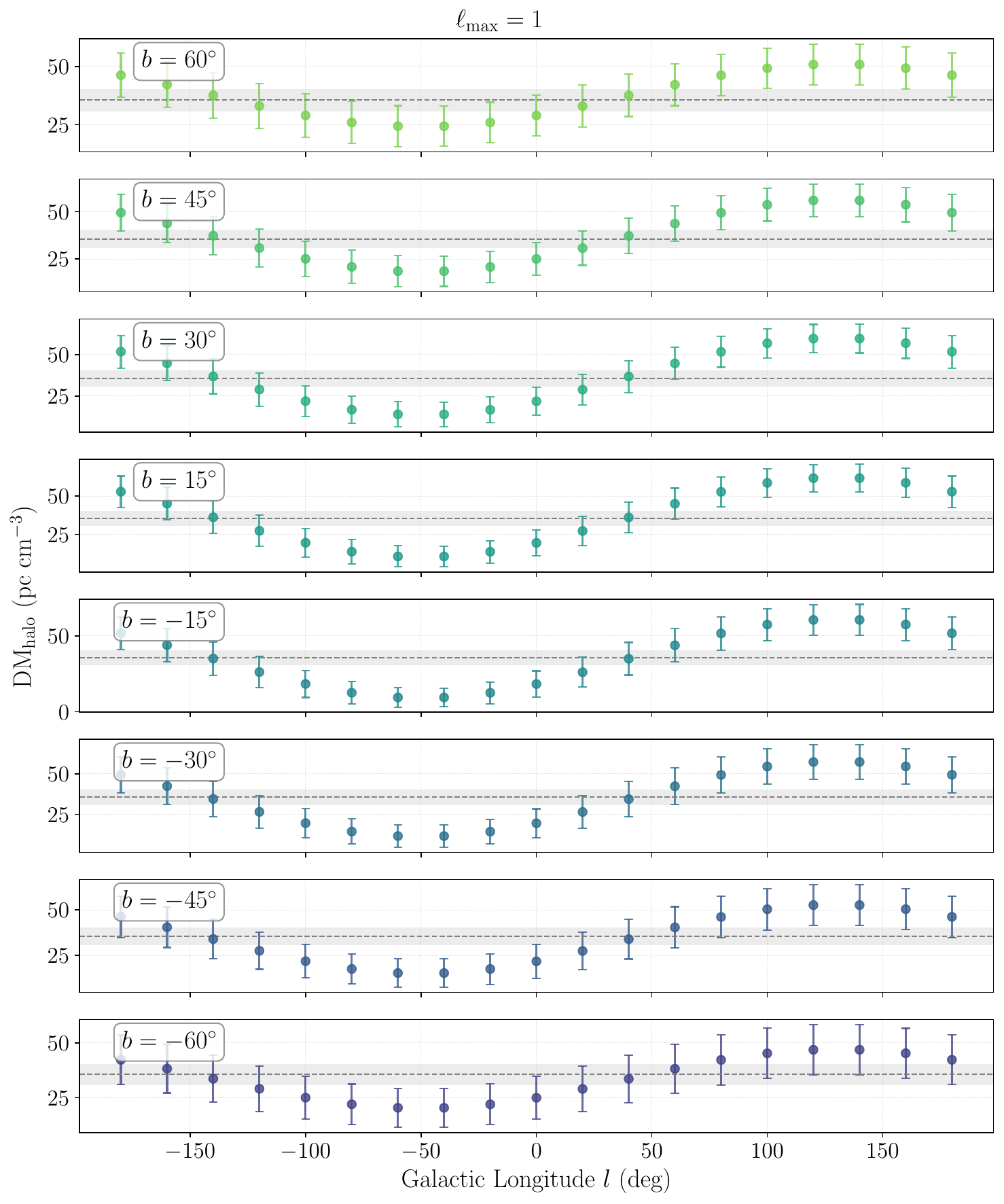}
	\includegraphics[width=0.45\textwidth]{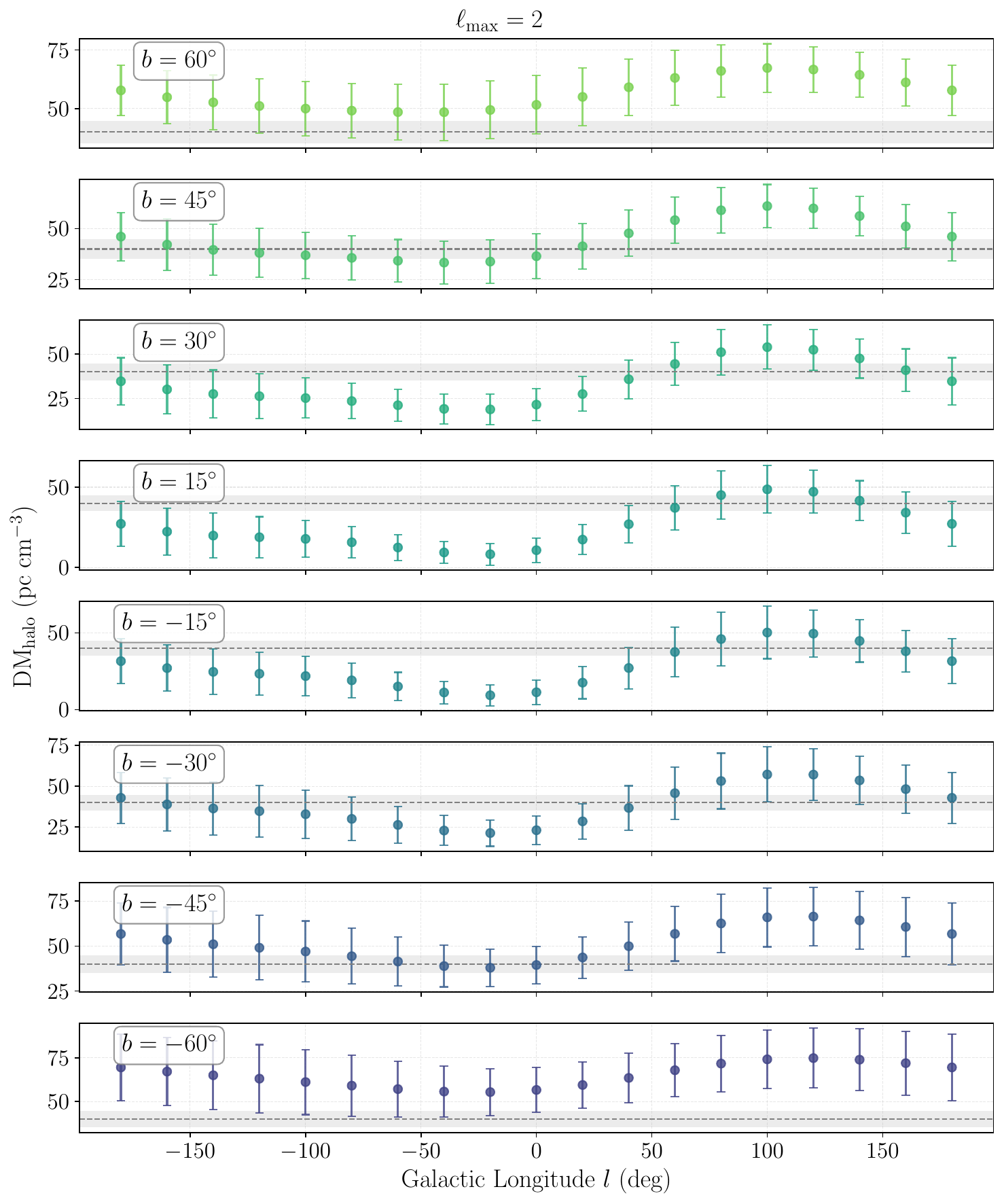}
	\caption{
		Predicted $\mathrm{DM_{halo}}$ as a function of Galactic longitude for different latitude slices.
		The left column shows the results for the $\ell_{\max}=1$ model, and the right column shows those for the $\ell_{\max}=2$ model. The gray shadows denote the all-sky mean values $\langle \mathrm{DM_{halo}} \rangle$ with 1$\sigma$ CL for both models.
		\label{fig:DM_lon}}
\end{figure}

\begin{figure}[bp]
	\centering
	\includegraphics[width=0.48\textwidth]{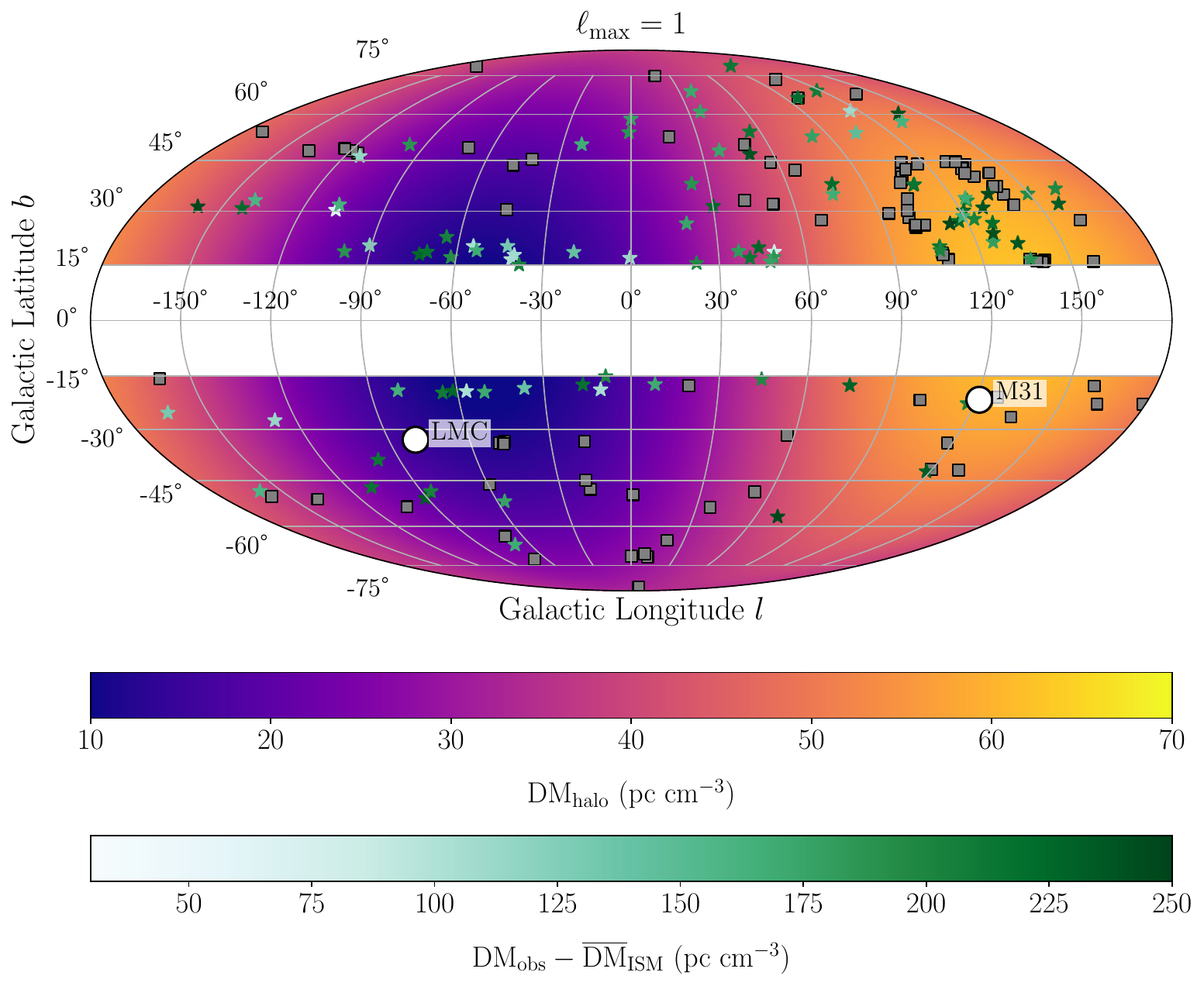}
	\includegraphics[width=0.48\textwidth]{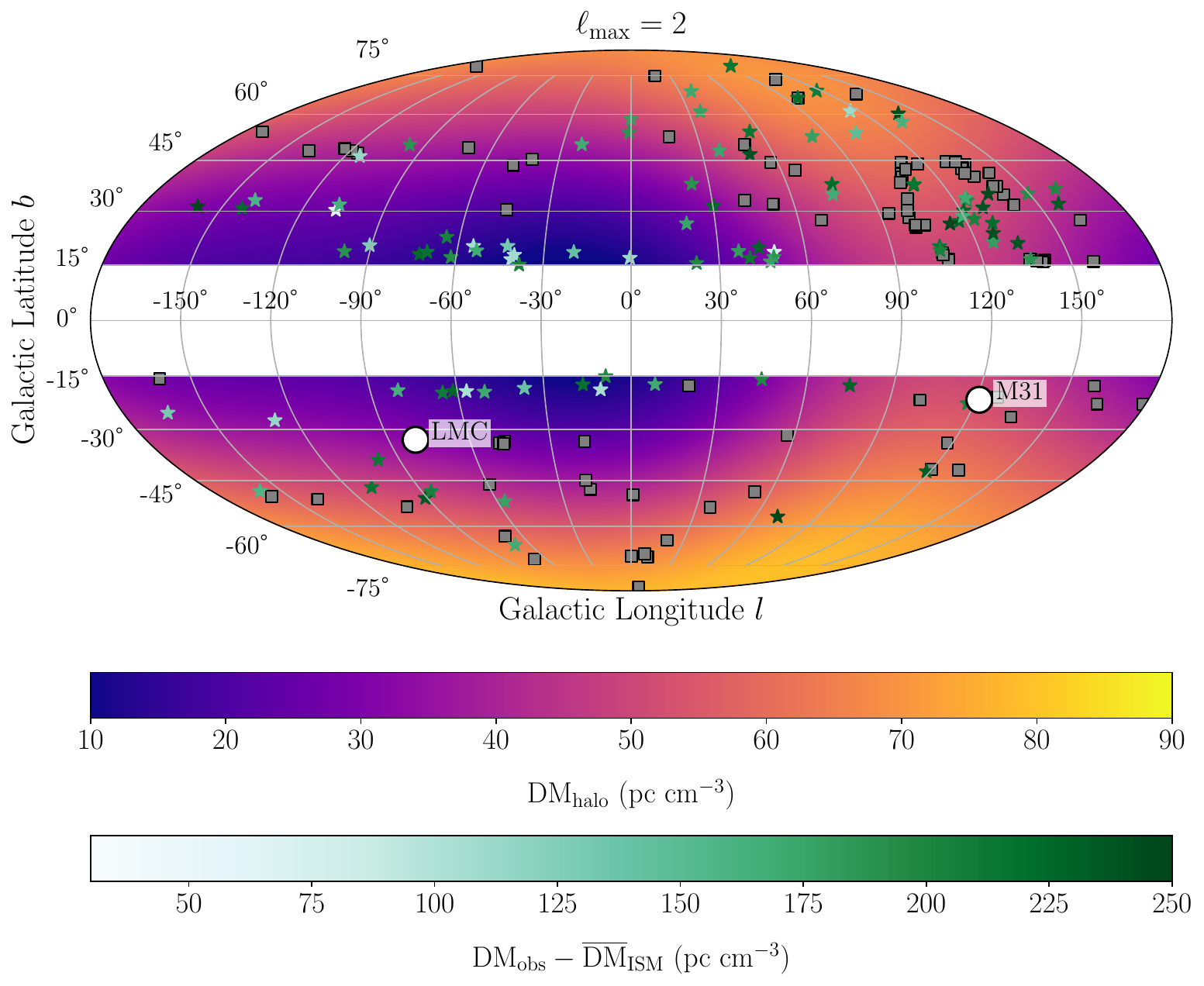}
	\caption{All-sky Mollweide projections of the $\mathrm{DM_{halo}}$ models expanded using spherical harmonics with $\ell_{\max}=1$ (left) and $\ell_{\max}=2$ (right). Gray squares denote localized FRBs, while stars represent unlocalized FRBs with $\mathrm{DM_{obs}}-\overline{\mathrm{DM}}_\mathrm{ISM}<250~\pccm$, where $\overline{\mathrm{DM}}_\mathrm{ISM}$ is calculated from the NE2001 model. The region within $|b|<15^\circ$ is masked out, and the directions of the LMC and M31 are also shown.
		\label{fig:map_ne2001}}
\end{figure}

Performing the Markov Chain Monte Carlo (MCMC) sampling, we obtain the posterior PDFs of $a_{\ell,m}$.
The one-dimensional (1D) posterior PDFs and two-dimensional (2D) confidence regions ($1\sigma$ and $2\sigma$) for $a_{\ell,m}$, derived using both the NE2001 and YMW16 models across different $\mathrm{DM_{halo}}$ models, are shown in Figures~\ref{fig:A_contour_ne2001} and~\ref{fig:A_contour_ymw16} of Appendix~\ref{sec:A1}. 
The corresponding mean values and the $1\sigma$ CLs are summarized in Table~\ref{tab:1}. 
As shown in this Table, the constraints on $a_{\ell,m}$ are largely insensitive to the choice of the Galactic ISM model (NE2001 or YMW16). Therefore, we will adopt the NE2001-based results as our main findings for the subsequent analysis.
Based on these constraints, we then calculate the $\mathrm{DM_{halo}}$ values at different Galactic directions using Equation~(\ref{eq:dm_harmonic_expansion}). 
The uncertainties of $\mathrm{DM_{halo}}$ are derived by error propagation as 
\begin{eqnarray}
	\sigma_\mathrm{DM_{halo}}^2=\sum_{i=1}^{N}\sum_{j=1}^{N} \left(\frac{\partial \mathrm{DM_{halo}}}{\partial \bm{a}_i }\right)\left(\frac{\partial \mathrm{DM_{halo}}}{\partial \bm{a}_j }\right) \bm{C}_{i,j},
\end{eqnarray}
where $\bm{a}=\left\{ a_{0,0}, a_{1,0}, a_{1,1}, \dots \right\}$ is a set of spherical harmonic coefficients $a_{\ell,m}$, $N$ is the number of $a_{\ell,m}$, and $\bm{C}$ is the covariance matrix of $a_{\ell,m}$.


For the $\ell_{\max}=1$ model based on NE2001, the coefficients $a_{1,1}$ and $a_{1,-1}$ deviate from zero by more than $2\sigma$, while $a_{1,0}$ remains consistent with zero within $1\sigma$.
An $a_{1,0}$ close to zero in our model indicates that the $\mathrm{DM_{halo}}$ is north-south symmetric.
To illustrate this characteristic, we plot the longitude-averaged $\mathrm{DM_{halo}}$, denoted as $\langle\mathrm{DM_{halo}}\rangle_\mathrm{lon}$, as a function of Galactic latitude:
\begin{eqnarray}
	\langle\mathrm{DM_{halo}}\rangle_\mathrm{lon}(b)=\frac{1}{2\pi}\int_{0}^{2\pi} \mathrm{DM_{halo}}(l,b)\ \mathrm{d}l.
\end{eqnarray}
The results are shown in Figure~\ref{fig:DM_lat}, together with the $\mathrm{DM_{obs}}-\overline{\mathrm{DM}}_\mathrm{ISM}$ values of both localized (gray squares) and unlocalized FRBs (green stars).
As we can see, the  $\langle\mathrm{DM_{halo}}\rangle_\mathrm{lon}$ remains nearly consistent across different latitudes.
The $\langle\mathrm{DM_{halo}}\rangle_\mathrm{lon}(-90^\circ)=33\pm11~\pccm$, which aligns will with $\langle\mathrm{DM_{halo}}\rangle_\mathrm{lon}(+90^\circ)=38\pm9~\pccm$.
However, the deviation of $a_{1,1}$ and $a_{1,-1}$ from zero suggests a dipole component in the Galactic longitude.
This trend can be found in the left panel of Figure~\ref{fig:DM_lon}, where $\mathrm{DM_{halo}}$ is plotted as a function of Galactic longitude at different Galactic latitude slices ($|b|$ from $15^\circ$ to $60^\circ$, in increments of $15^\circ$).
In this figure, we present the all-sky mean $\langle\mathrm{DM_{halo}}\rangle=36\pm5~\pccm$ value and its 1$\sigma$ range using a dashed line and gray shadow.
A clear excess of $\mathrm{DM_{halo}}$ is shown around $l= 130^\circ$, while a corresponding deficit appears around $l=310^\circ$. 
The magnitude of these deviations from the mean $\mathrm{DM_{halo}}$ increases significantly at lower Galactic latitudes.

This dipole anisotropic structure is clearly visible in the all-sky Mollweide projection of $\mathrm{DM_{halo}}$ shown in the left panel of Figure~\ref{fig:map_ne2001}, where the map is constructed using the mean values of the constraints on $a_{\ell,m}$. The $\mathrm{DM_{obs}}-\overline{\mathrm{DM}}_\mathrm{ISM}$ values of both localized FRBs (gray squares) and unlocalized FRBs (stars) are also plotted.
We find a peak value of $\mathrm{DM_{halo}}=63\pm9~\pccm$ toward the direction $(l=130^\circ, b=+5^\circ)$ and a minimum value of $\mathrm{DM_{halo}}=9\pm6~\pccm$ toward the direction $(l=310^\circ, b=-5^\circ)$. 
Relative to the all-sky mean $\langle\mathrm{DM_{halo}}\rangle=36\pm5~\pccm$, these represent deviations of about $2.6\sigma$ CL and $3.5\sigma$ CL from the mean value, respectively.
The 1$\sigma$ uncertainty of this dipole direction is about $28^\circ$.
It's worth noting that the mean $\mathrm{DM_{halo}}$ value in our model is in good agreement with the results obtained by \cite{2025AJ....169..330R}, who used an FRB at 50~Mpc to estimate an upper limit on $\mathrm{DM_{halo}}$ of either $28.7~\pccm$ or $47.3~\pccm$.
Moreover, our model predicts the $\mathrm{DM_{halo}}$ toward the LMC to be $\mathrm{DM_{halo}}=15\pm8~\pccm$.
Given that the NE2001 model predicts $\overline{\mathrm{DM}}_\mathrm{ISM} \approx 53~\pccm$ in the LMC direction and pulsar observations toward the LMC typically yield $\mathrm{DM_{MW}} \sim 70~\pccm$~\citep{2010ApJ...714..320A,2006ApJ...649..235M,2013MNRAS.433..138R}, our result also agrees with the pulsar-based estimate within $1\sigma$.

For the model with $\ell_{\max}=2$ based on the NE2001 model, the coefficient $a_{2,0}=52^{+20}_{-16}$ is significantly greater than zero at about $3.3\sigma$ CL, implying that the $\langle\mathrm{DM_{halo}}\rangle_\mathrm{lon}$ decreases toward lower Galactic latitudes. This trend is shown in Figure~\ref{fig:DM_lat}. 
The value of $\langle\mathrm{DM_{halo}}\rangle_\mathrm{lon}$ is $78\pm18~\pccm$ at $b=-90^\circ$ and is $68\pm12~\pccm$ at $b=+90^\circ$. 
In the southern hemisphere, the $\langle\mathrm{DM_{halo}}\rangle_\mathrm{lon}$ drops to $28\pm7~\pccm$ at $b=-15^\circ$, corresponding to a $2.6\sigma$ deviation from that at $b=-90^\circ$, while in the northern hemisphere it decreases to $26\pm6~\pccm$ at $b=+15^\circ$, representing a $3.1\sigma$ deviation from that at $b=+90^\circ$.

The coefficients  $a_{1,1}=20^{+13}_{-16}$ and $a_{1,-1}=-31^{+18}_{-17}$ in the $\ell_{\max}=2$ model continue to deviate from zero by more than $1\sigma$ CL, indicating a persistent $\mathrm{DM_{halo}}$ dipole component in the Galactic longitude.
The dependence of $\mathrm{DM_{halo}}$ on Galactic longitude at different latitudes is shown in the right panel of Figure~\ref{fig:DM_lon}. 
Similar to the $\ell_{\max}=1$ case, an excess of $\mathrm{DM_{halo}}$ persists across a wide range of latitudes and peaks near $l=100^\circ$.
Notably, however, in the $\ell_{\max}=2$ model, the most significant excess of $\mathrm{DM_{halo}}$ occurs at high latitudes ($b\gtrsim 45^\circ$), exceeding the all-sky mean value ($\langle\mathrm{DM_{halo}}\rangle=40\pm5~\pccm$) by more than $1\sigma$. 
This contrasts with the $\ell_{\max}=1$ model, where the excess is prominent at low latitudes.
This trend can also be clearly observed in the all-sky map (right panel of Figure~\ref{fig:map_ne2001}), indicating that both the $\ell_{\max}=1$ and $\ell_{\max}=2$ models reveal a longitude dependence of $\mathrm{DM_{halo}}$, while its latitude dependence remains less certain.

To evaluate which model is most favored by the current FRB data, we calculate the Akaike Information Criterion (AIC) from the likelihood values~\citep{1974ITAC...19..716A,Akaike1981}, defined as $\mathrm{AIC} = 2p - 2\ln \mathcal{L}$, where $p$ is the number of free parameters. Then, we compute the relative difference with respect to a reference model, defined as $\Delta \mathrm{AIC} = \mathrm{AIC}-\mathrm{AIC}_{\rm ref}$. Here the isotropic $\mathrm{DM_{halo}}$ model ($\ell_{\max}=0$) is adopted as the reference one. In this framework, $0<|\Delta \mathrm{AIC}|<2$ indicates that the models are indistinguishable, $4<|\Delta \mathrm{AIC}|<7$ provides moderate evidence against the model with the larger AIC, and $|\Delta \mathrm{AIC}|>10$ suggests strong evidence against it~\citep{Burnham2004}. 

Furthermore, to more comprehensively evaluate the performance of each model across its entire parameter space, we also calculate the Bayesian evidence ($\mathcal{Z}$) for each model~\citep{2007MNRAS.378...72T,2008ConPh..49...71T}.
This quantity, which is equal to the normalization constant of the parameter posterior distribution (Equation~(\ref{eq:post_L})), can be obtained by using $\mathcal{Z} = \int \mathcal{L}_\mathrm{FRB}(\mathrm{DM_{obs}}|\bm{\Theta})\times \Pi(\bm{\Theta})\; \mathrm{d}\bm{\Theta}$.
The relative preference between two competing models is quantified by the Bayes factor, defined as the ratio of their Bayesian evidences: $\mathcal{B}=\mathcal{Z}_0/\mathcal{Z}_1$. In our analysis, we always set the isotropic $\mathrm{DM_{halo}}$ model as the reference one ($\mathcal{Z}_0$). According to the Jeffreys scale~\citep{Jeffreys1961}, $|\ln \mathcal{B}|<1$ indicates an inconclusive preference between the two models, $1< |\ln \mathcal{B}|<2.5$ corresponds to weak evidence, $2.5<|\ln \mathcal{B}|< 5$ suggests moderate evidence, and $|\ln \mathcal{B}|>5$ implies strong evidence in favor of the model with a larger $\mathcal{Z}$.

The results are summarized in Table~\ref{tab:1}. 
When the NE2001 model is used, the AIC values for the models with $\ell_{\max}=0$, $\ell_{\max}=1$, and $\ell_{\max}=2$ are $1210$, $1200$, and $1200$, respectively, corresponding to strong evidence favoring both the $\ell_{\max}=1$ and $\ell_{\max}=2$ models. 
The logarithmic Bayes factor between the $\ell_{\max}=0$ and $\ell_{\max}=1$ models yields $\ln \mathcal{B}=-2$, also suggesting slight support for the $\ell_{\max}=1$ model.
However, the $\ln \mathcal{B}$ between the $\ell_{\max}=0$ and $\ell_{\max}=2$ models is $3.4$, indicating moderate evidence in favor of the isotropic $\mathrm{DM_{halo}}$ model.
These results indicate that the current FRB data slightly prefer the existence of a dipole structure in $\mathrm{DM_{halo}}$.
This conclusion holds even when the YMW16 model is used, as the $\Delta$AIC and $\ln \mathcal{B}$ values between the $\ell_{\max}=0$ and $\ell_{\max}=1$ models are $-7$ and $-1.9$, respectively.

\subsection{ Constraining $\mathrm{DM_{halo}}$ Models using Refined Sample }

Given that our full sample comprises FRBs from various experiments with differing sensitivities~\citep{2018Natur.562..386S}, we apply a stringent refinement to our sample to investigate whether the observed dipole anisotropy in $\mathrm{DM_{halo}}$ is an artifact of selection effects.
Since wideband RFI mitigation strategies are employed in CHIME, which preferentially remove signals from bright, low-DM events~\citep{2021ApJS..257...59C,2021ApJ...922...42R}, we exclude all FRBs detected by CHIME to avoid this instrumental selection effect.
Additionally, to prevent biases introduced by mixing data from different surveys, we exclude all unlocalized FRBs. We also discard known repeaters to avoid potential biases arising from different source types.
Then, the final sample comprises 62 localized FRBs (hereafter referred to as the `refined sample'), and the likelihood in Equation~(\ref{eq:post_L}) reduces from $\mathcal{L}_\mathrm{FRB}(\mathrm{DM_{obs}}|\bm{\Theta})$ to $\mathcal{L}_\mathrm{FRB}^\mathrm{loc}(\mathrm{DM_{obs}}|\bm{\Theta})$.

\begin{figure}[htbp]
	\centering
	\includegraphics[width=0.48\textwidth]{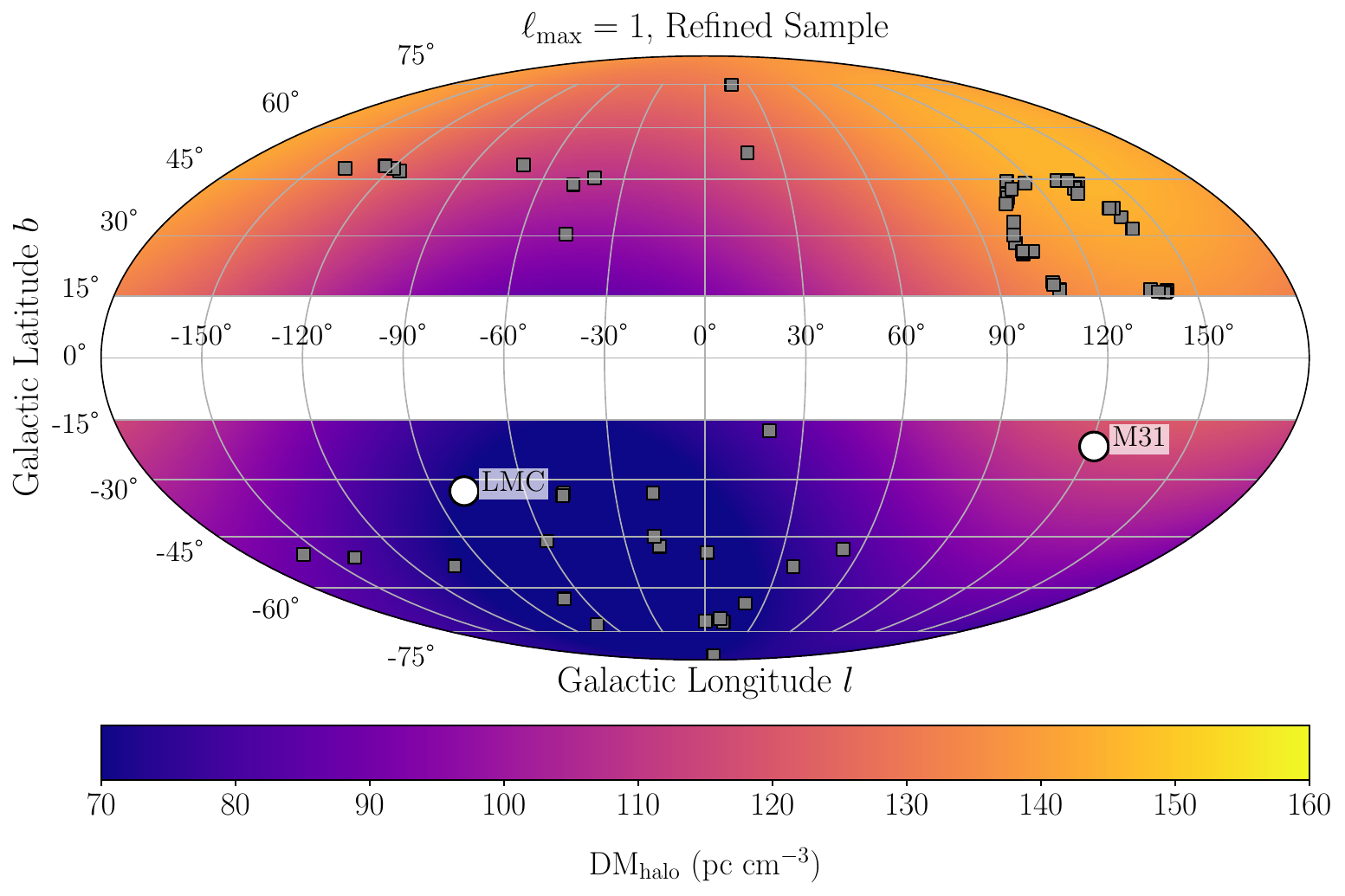}
	\caption{All-sky Mollweide projections of the $\mathrm{DM_{halo}}$ models expanded using spherical harmonics with $\ell_{\max}=1$. Only 62 localized FRBs are used (Gray squares).
		\label{fig:map_ne2001_F}}
\end{figure}

We consider only the $\ell_{\max}=0$ and $\ell_{\max}=1$ models, given that the latter was favored by both AIC and Bayesian evidence in the previous analysis.
Both the NE2001 and YMW16 models are also employed in this analysis, with the relevant results listed in Table~\ref{tab:2}.
Consistent with the results from the full sample (92 localized and 574 unlocalized FRBs), the choice of ISM model does not significantly affect the constraints on $a_{\ell,m}$.
Compared to the results obtained from the full sample, the uncertainties of $a_{\ell,m}$ increase significantly.
However, the coefficient $a_{1,-1}$ derived from both ISM models remains non-zero by more than $1\sigma$ CL.
In contrast to the full sample results, where the coefficient $a_{1,0}$ is consistent with zero, the refined sample reveals an $a_{1,0}$ value that deviates from zero by more than $1\sigma$. Consequently, the derived dipole direction shifts towards the Northern hemisphere compared to the full sample result.
The all-sky map of $\mathrm{DM_{halo}}$ based on the NE2001 model exhibits this dipole structure pointing toward $(l=141^\circ, b=+51^\circ)$, with a large $1\sigma$ uncertainty of approximately $44^\circ$ (Figure~\ref{fig:map_ne2001_F}). 
Comparing this to the dipole direction obtained from the full sample $(l=130^\circ, b=+5^\circ)$, we find that although there is an angular separation of $\sim 47^\circ$, the two directions are compatible within the $1\sigma$ CL, and the Galactic longitudes of the two dipoles are in good agreement.
Moreover, regardless of whether the NE2001 or YMW16 model is used, both AIC and Bayesian evidence favor the $\ell_{\max}=1$ model, yielding even stronger Bayesian evidence compared to the findings in the full sample.
These results indicate that the dipole anisotropic structure of $\mathrm{DM_{halo}}$ observed in the Galactic longitude direction is not caused by the CHIME sample selection, repeating sources, or unlocalized FRBs.

{%
	\renewcommand{\arraystretch}{1.5}%
	\begin{deluxetable*}{l|cccc|c|c|c}[tbp]
		\tablecaption{Constraints on $a_{\ell,m}$ of different $\mathrm{DM_{halo}}$ models using the refined sample  \label{tab:2}}
		\tablewidth{0pt}
		\tablehead{
			Model & $a_{0,0}$ & $a_{1,0}$ & $a_{1,1}$ & $a_{1,-1}$ & $-2\ln\mathcal{L}_{\max}$ & AIC\,($\ln \mathcal{Z}$) & $\Delta\mathrm{AIC}\,(\ln \mathcal{B})$
		}
		\startdata
		\hline
		\multicolumn{8}{c}{ 62 localized FRBs (NE2001)} \\
		\hline
		$\ell_{\max}=0$ & $306^{+41}_{-30}$ & - & - & - & 797 & 807 ($-401.9$) & 0 (0)  \\
		$\ell_{\max}=1$ & $369^{+49}_{-43}$ & $66^{+40}_{-39}$ & $41^{+67}_{-55}$ & $-34^{+31}_{-30}$ & 780 & 796 ($-396.9$) & $-11$ ($-5.0$) \\
		\hline
		\hline
		\multicolumn{8}{c}{ 62 localized FRBs (YMW16)} \\
		\hline
		$\ell_{\max}=0$ & $309^{+37}_{-28}$ & - & - & - & 800 & 810 ($-403.6$) & 0 (0)  \\
		$\ell_{\max}=1$ & $378^{+47}_{-37}$ & $74^{+46}_{-36}$ & $25^{+61}_{-66}$ & $-34^{+30}_{-30}$ & 784 & 800 ($-398.8$) & $-10$ ($-4.8$) \\
		\enddata
		\tablecomments{All results are given as mean values with 1$\sigma$ uncertainties. 
			Here $-2\ln\mathcal{L}_{\max}$ represents $-2$ times the maximum log-likelihood value. 
			$\Delta \mathrm{AIC=AIC-AIC_{ref}}$, and we set the model with $\ell_{\max}=0$ as the reference one. $\ln\mathcal{Z}$ is the logarithmic Bayesian evidence and $\ln \mathcal{B}= \ln\mathcal{Z}_0-\ln\mathcal{Z}_1$, where $\ln\mathcal{Z}_0$ and $\ln\mathcal{Z}_1$ are the logarithmic Bayesian evidences for the reference isotropic $\mathrm{DM_{halo}}$ model ($\ell_{\max}=0$) and a competing alternative model, respectively.
		}
	\end{deluxetable*}
}

It is worth noting that if the Local Group contains a significant amount of intragroup medium (IGrM), it would also contribute to the DM of FRBs. Modeling the Local Group IGrM with a dark matter halo mass of $10^{12.5}M_\odot$ suggests that it could yield a significant DM excess toward M31~\citep{2019MNRAS.485..648P}.
The Local Group IGrM has also been studied by \cite{2021ApJ...907...14Q}, who found a large-scale hot-gas content toward M31 through analyses of X-ray O\,{\footnotesize VII} and O\,{\footnotesize VIII} emission lines and Sunyaev-Zel'dovich maps, suggesting the presence of a possible Local Hot Bridge connecting the MW and M31. 
\cite{2014MNRAS.441.2593N} also reported a significant excess of gas between the MW and M31, by using a constrained cosmological simulation of the Local Group. 
These results suggest that gas within the Local Group may contribute considerably to the DM toward M31.
In our results, we find that the direction of DM excess is consistent with the position of M31 ($l=121.2^\circ, b=-21.6^\circ$).
The angular separation between M31 and the dipole direction of the $\ell_{\max}=1$ model is approximately $28^\circ$ for the full sample and $75^\circ$ for the refined sample (see Figures \ref{fig:map_ne2001} and \ref{fig:map_ne2001_F}). 
These separations fall within the $1\sigma$ and $2\sigma$ CLs, respectively. 
Consequently, the Local Group IGrM may contribute to the detected $\mathrm{DM_{halo}}$ excess in the $\ell_{\max}=1$ model.

On the other hand, eROSITA has recently revealed a pair of soft-X-ray-emitting bubbles extending about 14 kpc above and below the Galactic center, known as the eROSITA bubbles~\citep{2020Natur.588..227P}. These bubbles exhibit significant asymmetries. Hydrodynamic simulations indicate that such features can be explained by a dynamic CGM wind model~\citep{2023NatCo..14..781M}, in which a wind from the east-north direction in Galactic coordinates crosses the northern halo at a velocity of $\sim200~\mathrm{km~s^{-1}}$. 
Compression of gas on the windward side could naturally lead to an enhanced density, which may also explain the excess $\mathrm{DM_{halo}}$ we find in the $\ell_{\max}=1$ model. 
Future FRB observations with expanded sample sizes and improved sky coverage are essential to verify these hypotheses.

\section{Conclusion}\label{sec:conclusion}

We propose a data-driven approach to reconstruct the all-sky $\mathrm{DM_{halo}}$ distribution by expanding it in spherical harmonics. 
Using the full sample of 92 localized and 574 unlocalized non-repeating FRBs at $|b|>15^\circ$, we constrain the expansion coefficients $a_{\ell,m}$ for $\mathrm{DM_{halo}}$ models with degrees $\ell_{\max}=0,1,2$. 
For the model of $\ell_{\max}=1$ based on the NE2001 model, we find a significant dipole anisotropic structure in $\mathrm{DM_{halo}}$ pointing toward $(l=130^\circ, b=+5^\circ)$, with a $1\sigma$ uncertainty of $\sim 28^\circ$. 
Along this direction, the inferred $\mathrm{DM_{halo}}=63\pm9~\mathrm{pc~cm^{-3}}$ exceeds the all-sky mean value of $\langle \mathrm{DM_{halo}} \rangle = 36\pm5~\mathrm{pc~cm^{-3}}$ by approximately $2.6\sigma$. 
When extending the analysis to the $\ell_{\max}=2$ model, the excess near $l\approx 100^\circ$ persists, although the most significant excess appears at high latitudes ($b\gtrsim 45^\circ$). 
These results are not significantly affected by the choice of Galactic ISM models.
Even when the full sample is refined to only 62 localized FRBs (excluding CHIME detections, repeaters, and unlocalized events), the dipole anisotropic structure persists, and its direction is consistent with the full sample result within $1\sigma$ CL.

To assess the relative statistical significance of these models, we calculate both the AIC and the Bayesian evidence for each model. 
For the full sample, we find $\Delta \mathrm{AIC} = -10$ for the $\ell_{\max}=1$ model compared to the isotropic $\mathrm{DM_{halo}}$ model, strongly favoring the existence of a dipole structure.
A similar strong preference is also obtained for the $\ell_{\max}=2$ model using the AIC.
In contrast, the Bayesian evidence implies a more conservative conclusion.
The logarithmic Bayes factor for the isotropic model relative to the $\ell_{\max}=1$ model is $\ln \mathcal{B} = -2$, suggesting a weak preference for the dipole structure.
For the $\ell_{\max}=2$ model, we find $\ln \mathcal{B} = 3.4$, providing moderate evidence against the $\ell_{\max}=2$ model.

Considering both the AIC and Bayesian evidence, we find that the current FRB data slightly favor the existence of a dipole structure in $\mathrm{DM_{halo}}$.
If such a dipole is real and not a result of statistical fluctuations or systematic error, it might be caused by the IGrM in Local Group or a CGM wind, which requires detailed investigation in future work.
Future samples of FRBs with broader sky coverage and more localized sources will be essential to confirm or refute this dipole feature and to clarify its physical origin.

\begin{acknowledgements}
This work is supported by
the National Natural Science Foundation of China (grant Nos. 12321003, 12422307, 12373053,
and 12275080), the CAS Project for Young Scientists in Basic Research (grant No. YSBR-063), 
and China Postdoctoral Science Foundation (grant No. 2025M783235).
\end{acknowledgements}


	

 \appendix
 \section{MCMC inference of spherical harmonic coefficients} \label{sec:A1}
The 1D posterior PDFs and 2D confidence regions (with 1-2$\sigma$) for $a_{\ell,m}$ in different $\mathrm{DM_{halo}}$ models and ISM models (NE2001 and YMW16) are shown here. All results are derived from the full sample (92 localized and 574 unlocalized FRBs).

\begin{figure}[htbp]
	\centering
	\includegraphics[width=0.8\textwidth]{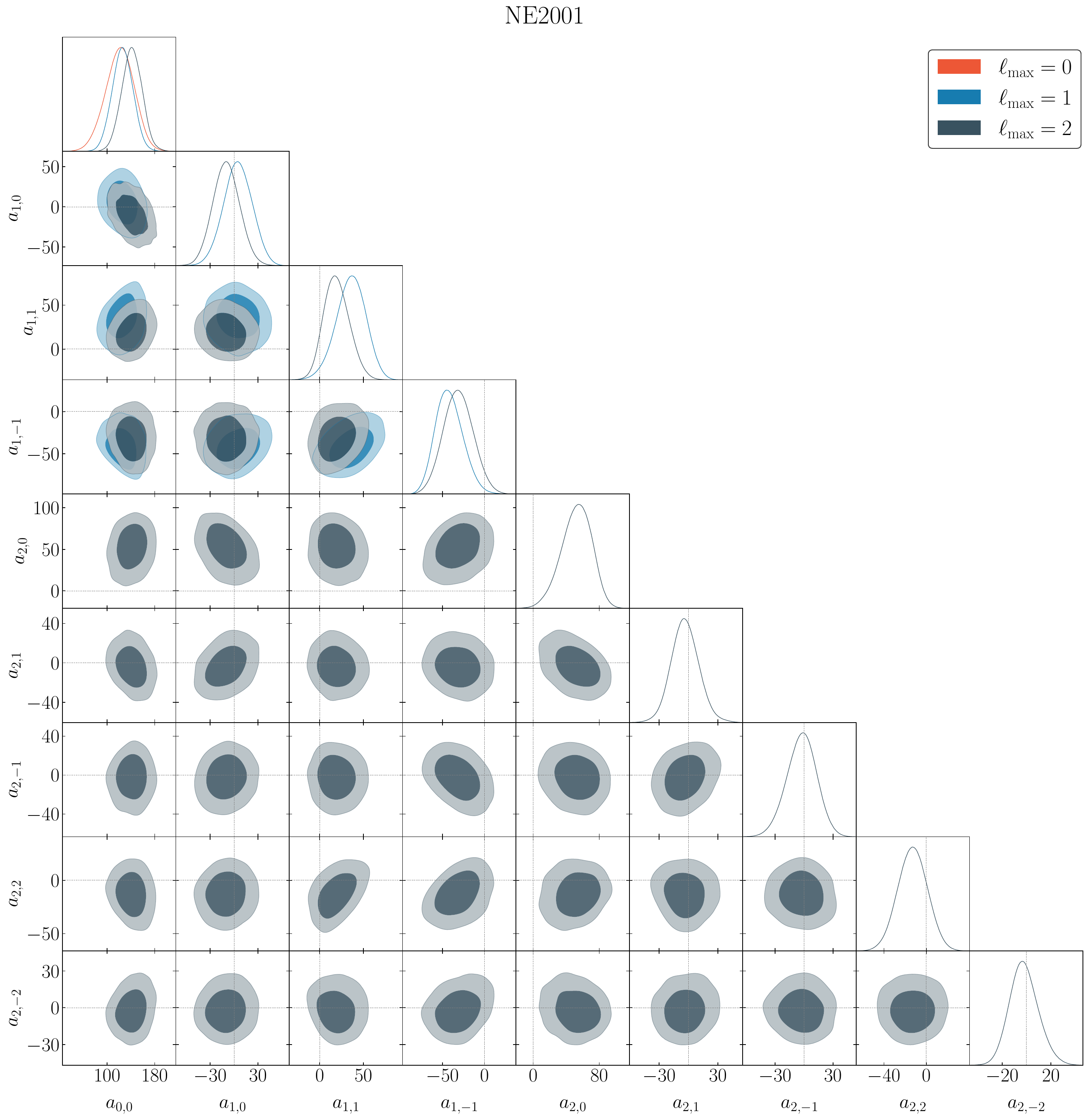}
	\caption{
		1D posterior PDFs and 2D confidence regions (with 1–2$\sigma$ contours) for $a_{\ell,m}$, shown in red for $\ell_{\max}=0$, blue for $\ell_{\max}=1$, and gray for $\ell_{\max}=2$. All results are derived from the full sample based on the NE2001 model.
		\label{fig:A_contour_ne2001}}
\end{figure}

\begin{figure}[htbp]
	\centering
	\includegraphics[width=0.8\textwidth]{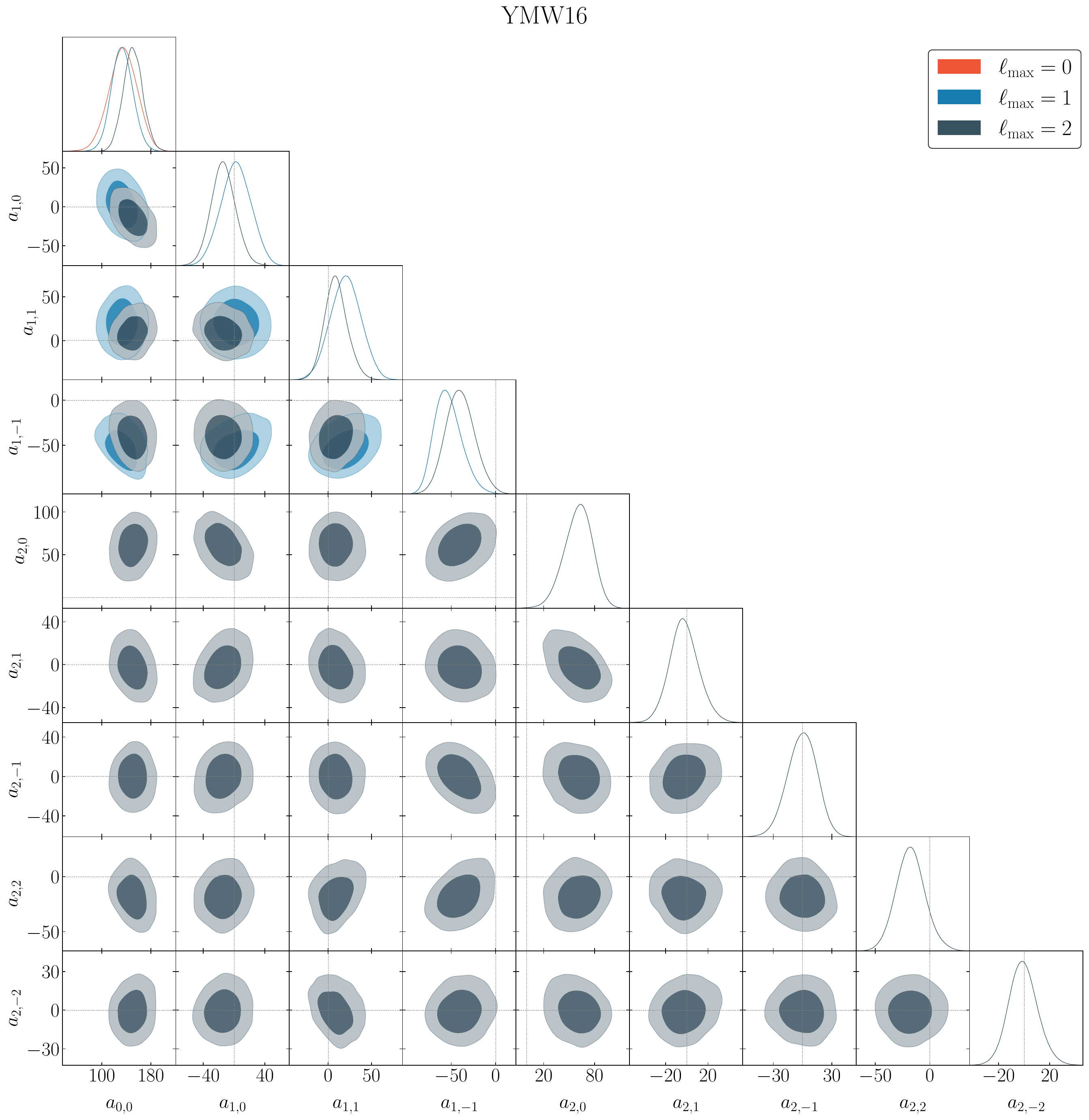}
	\caption{
		Similar to Figure~\ref{fig:A_contour_ne2001}, except the ISM model is the YMW16 model.
		\label{fig:A_contour_ymw16}}
\end{figure}

\bibliography{ref}{}
\bibliographystyle{aasjournalv7}

\end{document}